\documentclass[12pt]{article}
\usepackage{latexsym}
\usepackage{amsfonts,color}
\usepackage{amssymb,amsmath,amsthm}
\usepackage{epsfig}
\usepackage{graphicx}

\newcommand{\rem}[1]{}
\newcommand{\remfigure}[1]{#1}
\newcommand{\beq}{\begin{equation}}
\newcommand{\eeq}{\end{equation}}
\newcommand{\bea}{\begin{eqnarray}}
\newcommand{\eea}{\end{eqnarray}}

\begin{document}

\title{A class of equations with peakon and pulson solutions (with an Appendix 
by Harry Braden and John Byatt-Smith)} 
\author{Darryl D. Holm\thanks{Computer and Computational Science Division, 
Los Alamos National Laboratory, MS D413,  
Los Alamos, 
NM 87545. E-mail:
dholm@lanl.gov}
\thanks{  
Mathematics Department, Imperial College London, SW7 2AZ, U.K.  
E-mail: 
d.holm@imperial.ac.uk} 
and Andrew N.W. Hone\thanks{Institute of Mathematics and Statistics, 
University of Kent,   
Canterbury CT2 7NF, U.K.  
E-mail: 
A.N.W.Hone@kent.ac.uk}}  

\maketitle 
\bigskip

\centerline
{\it To Francesco Calogero on the Occasion of his 70th Birthday}
  
\begin{abstract}

\noindent
We consider a family of integro-differential equations 
depending upon a parameter $b$ as well as a symmetric integral kernel $g(x)$. 
When $b=2$ and $g$ is the peakon kernel  
(i.e. $g(x)=\exp(-|x|)$ up to rescaling)  the dispersionless Camassa-Holm   
equation results, while the Degasperis-Procesi equation is obtained 
from the peakon kernel with $b=3$. Although these two cases are integrable, 
generically the corresponding integro-PDE is non-integrable. However, 
for $b=2$ the family restricts to the pulson family of Fringer \& Holm, 
which is Hamiltonian and numerically displays elastic scattering of pulses. 
On the other hand, for arbitrary $b$ it is still possible to construct 
a nonlocal Hamiltonian structure provided that $g$ is the peakon kernel 
or one of its degenerations: we present a proof of this fact using an 
associated functional equation for the skew-symmetric antiderivative of $g$. 
The nonlocal bracket reduces to a non-canonical Poisson bracket for 
the peakon dynamical system, for any value of $b\neq 1$.   
\end{abstract}

\section{Introduction}

Here we will consider a class of integro-partial differential 
equations (IPDEs) of the form  
\beq 
m_t+um_x+bu_xm=0, \qquad u=g*m, 
\label{ipde} 
\eeq     
where $g\, *$ denotes convolution with a symmetric integral kernel $g(x)$ 
defined on the real line $\mathbb{R}$: 
$$ 
u(x)=g*m(x)=\int_{-\infty}^\infty g(x-y)m(y)\, dy. 
$$ 
There are some distinguished special cases of this equation, the first 
being the dispersionless version of the integrable Camassa-Holm equation,
\beq
u_t-u_{xxt}+3uu_x=2u_xu_{xx}+uu_{xxx},
\label{eq:caholm}
\eeq
which is the $b=2$ case of (\ref{ipde}) when the kernel is chosen to 
be the Green's function for the Helmholtz operator on the line: 
\beq \label{peaker} 
g(x)=\frac{1}{2}e^{-|x|}, \qquad m=u-u_{xx}. 
\eeq 
With the inclusion of additional linear dispersion terms the equation
(\ref{eq:caholm}) was derived
as an approximation to the incompressible Euler
equations, and found to be completely integrable with a
Lax pair and associated bi-Hamiltonian structure \cite{ch,ch2} 
(for another derivation, see \cite{johnson}). 
In fact the Camassa-Holm equation
fits into the framework of
hereditary symmetries and recursion operators described
earlier by Fokas and Fuchssteiner \cite{ff}.
A remarkable discovery of \cite{ch,ch2} was that
in the dispersionless limit (\ref{eq:caholm}), the
solitons of the Camassa-Holm equation are peakons,
given by a superposition of an arbitrary number of peaks,
\beq
u(x,t)=\frac{1}{2}\sum_{j=1}^Np_j(t)e^{-|x-q_j(t)|},
\label{eq:Npeakon}
\eeq
(See also \cite{beals, constantin, constantin3}.) 
The $N$-peakon solutions (\ref{eq:Npeakon}) are weak
solutions with  discontinuous first derivatives at the
positions $q_j$ of the peaks; $q_j$, $p_j$ are
canonical coordinates and momenta in an
integrable finite-dimensional Hamiltonian system; the interpretation 
of weak solutions is discussed further in \cite{constantin2}. 
The $N$-peakon interaction was obtained in \cite{beals2}, 
with the explicit peakon-antipeakon interaction being given in 
\cite{beals3}. The stability  of peakons in the Camassa-Holm 
equation was proved in \cite{constantin4, constantin5}.  

In a recent application of the method of
asymptotic integrability to a
many-parameter family
of third order equations with dispersion, 
Degasperis and Procesi \cite{dega}
found that only three equations
passed the test up to third order in the asymptotic expansion,
namely KdV, Camassa-Holm and one new equation. After
rescaling and applying a Galilean transformation, the new
equation may be written in dispersionless form as
\beq
u_t-u_{xxt}+4uu_x=3u_xu_{xx}+uu_{xxx}. \label{eq:tdnodisp}
\eeq
Henceforth we refer to (\ref{eq:tdnodisp}) as the
Degasperis-Procesi equation, and observe that it
is the $b=3$ case of the equation (\ref{ipde}) with 
$g(x)=e^{-|x|}/2$. 
In a recent article \cite{needs} we proved the integrability
of the Degasperis-Procesi equation by constructing a Lax
pair, and derived two infinite sequences of conservation laws.

\remfigure{
\begin{figure}[ht!]
\centerline{
\scalebox{0.3}{\includegraphics{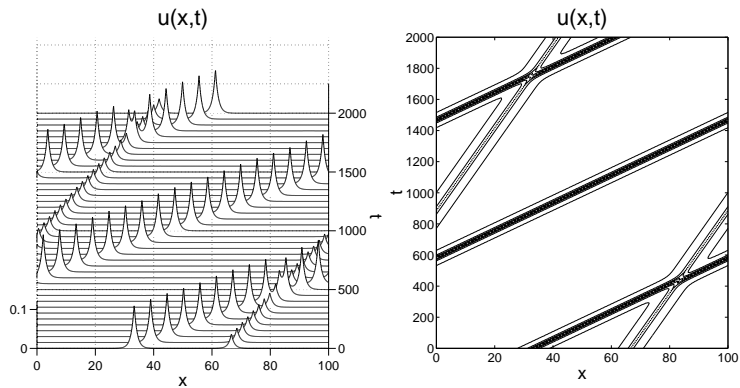}}
}
\caption{ {\it The elastic scattering of two peakons in the
integrable Degasperis-Procesi equation
on a periodic domain. This corresponds to $b=3$ and $g(x)=e^{-|x|}/2$ 
in (\ref{ipde}).} }
\label{peakons}
\end{figure}
}

In order to better understand the common properties of the two integrable 
equations (\ref{eq:caholm}) and         
(\ref{eq:tdnodisp}), we have found it convenient to consider the following
one-parameter family of partial differential equations
(PDEs):
\beq
u_t-u_{xxt}+(b+1)uu_x=bu_xu_{xx}+uu_{xxx}.   \label{eq:bfamily}
\eeq
Since it arises from (\ref{ipde}) when the peakon kernel (\ref{peaker}) 
is chosen, we refer to (\ref{eq:bfamily}) as the peakon $b$-family 
of PDEs. Remarkably, every member of this family 
admits
multi-peakon solutions (\ref{eq:Npeakon}), with a non-canonical
Hamiltonian structure for the dynamics of $q_j$, $p_j$
found in \cite{galli}. The two-body dynamics is integrable
for any $b$ (see Figure \ref{peakons}), and for all values of $b>1$
the peakons appear to
be numerically stable and dominate the initial value
problem \cite{hs}. Furthermore, with the addition of linear 
dispersion ($u_x$ and $u_{xxx}$ terms) it has been shown 
that not only the Camassa-Holm equation \cite{dgh1} but also 
the whole $b$-family (\ref{eq:bfamily}) (apart from $b=-1$) 
belong to a class of asymptotically equivalent shallow 
water wave equations \cite{dgh2}.  

In \cite{galli}  we presented a Lagrangian formulation 
for the $b$-family (\ref{eq:bfamily}), with a Legendre transformation 
leading to the Hamiltonian structure  
\beq 
m_t=\hat{B}\frac{\delta H}{\delta m}, 
\label{hamstr} 
\eeq 
where for $b\neq 1$  
\beq
\hat{B}=-(bm\partial_x+m_x)(\partial_x-\partial_x^3)^{-1}
(bm\partial_x+(b-1)m_x), \qquad H=\frac{1}{b-1}\int m\, dx.   
\label{eq:hamop}
\eeq
(In the case $b=1$ it is necessary to take $H=\int m\log m\, dx$ 
instead.) 
The skew-symmetric operator $\hat{B}$ in (\ref{eq:hamop}) 
was first obtained in \cite{needs} for the case $b=3$, as 
a second Hamiltonian structure for the Degasperis-Procesi equation 
(\ref{eq:tdnodisp}). However, a proof of the Jacobi identity for 
this operator was lacking until the work of one of us with 
Wang \cite{wanghone}, where the trivector formalism of Olver 
\cite{olver} was 
applied to this problem. The first two Hamiltonian 
structures of (\ref{eq:tdnodisp}) are 
\beq
B_0=-\partial_x(1-\partial_x^2)(4-\partial_x^2),
\qquad B_1=\hat{B}|_{b=3}=-9m^{2/3}\partial_xm^{1/3} 
(\partial_x-\partial_x^3)^{-1}m^{1/3}\partial_xm^{2/3},
\label{eq:tdops}
\eeq
while for the Camassa-Holm equation (\ref{eq:caholm})  
the bi-Hamiltonian structure arising from the Lax pair in 
 \cite{ch, ch2} is 
$$
B_0=-\partial_x(1-\partial_x^2), \qquad
B_1=-(m\partial_x+\partial_x m),
$$
and applying the recursion operator $R=B_1 B_0^{-1}$ to
$B_1$ gives the nonlocal operator 
$$
B_2\equiv B_1 B_0^{-1}B_1=\hat{B}|_{b=2}.
$$
Moreover, it is proved in \cite{wanghone} that $b=2,3$ are the only 
parameter values for which the operator $\hat{B}$ given by 
(\ref{eq:hamop}) is compatible with another operator with constant 
coefficients. The integrable cases $b=2,3$ can also be isolated by the 
perturbative symmetry approach \cite{mik}, by the Wahlquist-Estabrook 
prolongation algebra method \cite{wanghone}, or by using a reciprocal 
transformation and then applying Painlev\'e analysis \cite{pain}.  

In the special case of $b=2$ with an arbitrary 
symmetric kernel $g(x)$, the 
equation (\ref{ipde}) has the Lie-Poisson structure 
\beq 
m_t=\{ m,\tilde{H} \}_{LP}:=
 -(m\partial_x+\partial_x m)\frac{\delta \tilde{H}}{\delta m} 
=-ad^*_u m,
\label{liepoisson}
\eeq
with the quadratic Hamiltonian 
$$ 
\tilde{H}=\frac{1}{2}\int m\,g*m \, dx, \qquad 
\frac{\delta \tilde{H}}{\delta m}=u. 
$$ 
Equivalently the equation (\ref{liepoisson}) corresponds to 
the Euler-Poincare equation for geodesic motion on the 
diffeomorphism group of the real line, 
assuming that $g$ (the co-metric) is the 
Green's function for a positive definite operator (see \cite{fringer}  
for more details, and see \cite{homa} 
for an extension to higher 
dimensions). In the periodic case the Camassa-Holm equation is 
the equation for geodesics on the diffeomorphism group of the circle 
\cite{misiolek}, leading to a proof of the 
corresponding least action principle 
\cite{constantin6, constantin7}.   

In the case $b=2$ given by (\ref{liepoisson}) 
the equation admits special solutions 
in the form of a finite superposition of pulsons whose shape is 
determined by $g$, that is      
\beq 
u=\sum_{j=1}^N p_j(t) \, g(x-q_j(t)), \qquad 
m=\sum_{j=1}^N p_j(t) \, \delta (x-q_j(t)). \label{pulsons} 
\eeq 
The finite-dimensional dynamical system for the motion of $N$ pulsons 
on the line is a canonical Hamiltonian system for geodesic motion 
on an $N$-dimensional manifold with coordinates $q_j$, $j=1,\ldots ,N$ 
and co-metric $g(q_j-q_k)$, derived from the Hamiltonian 
\beq 
\label{geoham} 
\tilde{h}=\frac{1}{2}\sum_{j,k} 
p_jp_k\, g(q_j-q_k). 
\eeq 
The canonical Poisson bracket 
\beq 
\label{canon} 
\{ q_j, p_k\}=\delta_{jk} \qquad \{ q_j, q_k\}=0=\{ p_j, p_k\}   
\eeq 
can be obtained by substituting the ansatz (\ref{pulsons}) for 
$m$  
into the Lie-Poisson bracket 
$$\{ m(x), m(y) \}_{LP}$$ defined by the 
Hamiltonian operator $-(m\partial_x+\partial_x m)$, and similarly 
$\tilde{h}$ in (\ref{geoham}) is derived from $\tilde{H}$ in 
(\ref{liepoisson}) by substitution.  
Furthermore, although the equation (\ref{liepoisson}) 
is generically non-integrable, numerical studies in 
\cite{fringer} provide strong evidence that the pulson solutions 
(\ref{pulsons}) are stable and are produced by arbitrary smooth 
initial data, just as for  the integrable case of peakons in 
the Camassa-Holm equation. We have proved non-integrability 
for a particular fifth order equation in this class \cite{hoho}, although 
numerically the pulsons appear to scatter elastically (see figure 2).    

\remfigure{
\begin{figure}[ht!]
\centerline{
\scalebox{0.45}{\includegraphics{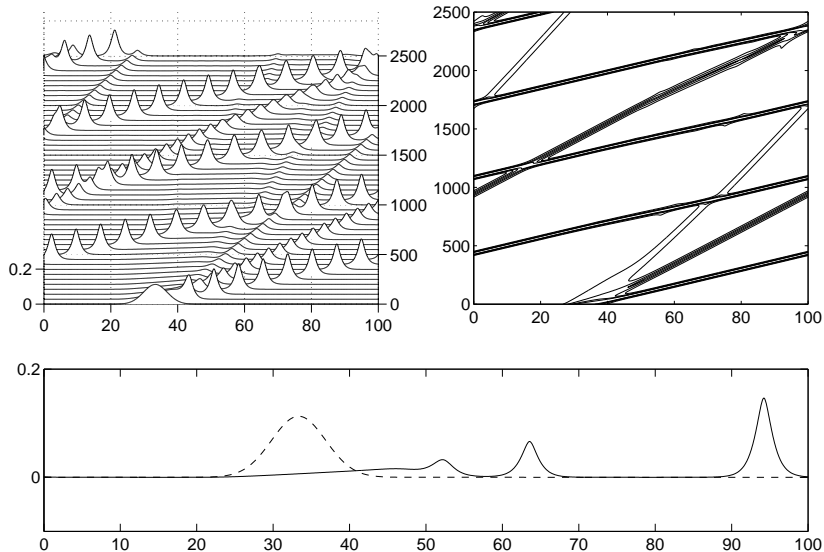}}
}
\caption{Pulson solutions (\ref{pulsons}) of
equation
  (\ref{liepoisson}) with $g(x)=2e^{-|x|}-e^{-2|x|}.$ 
emerge from a Gaussian of unit
  area and width $\sigma = 5$ centered about $x = 33$
on a periodic domain
  of length $L = 100$.  The fastest pulson crosses the
domain four times
  and collides elastically with the slower ones.}
\label{Gauss_InitialCond-waterfall/contour_fig}
\end{figure}}

One might wonder why we have chosen to discuss this subject in a 
birthday volume in honour of Francesco Calogero. In fact 
Calogero has already studied the problem of finding 
Lax pairs for integrable finite-dimensional systems with 
geodesic Hamiltonians of the form (\ref{geoham}), and obtained 
functional equations in terms of the function $g(x)$ and the 
components of the Lax matrix (see 
\cite{calogero} and references). Here we are concerned with a
slightly different problem. It turns out that the pulson ansatz 
(\ref{pulsons}) works equally well for the whole family of 
IPDEs (\ref{ipde}), and leads to the dynamical system 
\beq
\frac{dq_j}{dt}   =  
\sum_{k=1}^N p_kg(q_j-q_k),  \qquad  
\frac{dp_j}{dt}   =  
-(b-1)\sum_{k=1}^N p_jp_kg'(q_j-q_k).
\label{eq:peakdyn}
\eeq
Clearly this takes the canonical Hamiltonian form, 
with the Hamiltonian $\tilde{h}$ given by (\ref{geoham}),  
only for the case $b=2$. However, in the particular case $b=3$ and 
$g(x)=e^{-|x|}/2$ that corresponds to the integrable Degasperis-Procesi 
equation, we have derived a matrix Lax pair for the system 
(\ref{eq:peakdyn}), and therefore in this case 
the system must be integrable with respect to some Poisson structure. 
Requiring the existence of a certain type of  
Poisson structure for 
the IPDEs (\ref{ipde}) leads us to the functional equation 
\begin{equation}
G'(\alpha)(G(\beta)+G(\gamma))+
G'(\beta)(G(\gamma)+G(\alpha))
+G'(\gamma)(G(\alpha)+G(\beta))=0 \,  
\label{eq:fn1}
\end{equation}
when $\alpha+\beta+\gamma=0$,  for the antisymmetric kernel 
\beq \label{anti}  
G(x)=\int_0^x g(t) dt.
\eeq 
Apart from some limiting cases, the general odd solution of 
(\ref{eq:fn1}) turns out to be 
$$ 
G(x)=A\, \mathrm{sgn}(x)\Big(1-e^{-B|x|}\Big), 
\qquad \mathrm{whence} \qquad g(x)=ABe^{-B|x|} 
\quad (A,B \, \mathrm{constant}). 
$$ 
(The derivation of the general solution of 
(\ref{eq:fn1})  
is presented in an  Appendix provided by H.W. Braden and 
J.G. Byatt-Smith.) Thus 
the peakon $b$-family (\ref{eq:bfamily}) is picked out of the 
class (\ref{ipde}) by the functional equation (\ref{eq:fn1}). 
Moreover, for any $b$ the corresponding Hamiltonian operator, given by 
(\ref{eq:hamop}), reduces to a Poisson bracket for  
the $N$-peakon dynamical system, first presented without 
proof in \cite{galli}. 
In the next section we derive the functional equation from 
the Jacobi identity for the operator (\ref{eq:hamop}), and then 
proceed to show how this leads  to a suitable bracket for 
the peakons.    

\paragraph{Historical Note.} 
This paper is dedicated to our friend,
Francesco Calogero. This work parallels some of his earlier considerations
on geodesic flows and the determination of integrability for dynamical 
systems by using functional equations. Calogero's fundamental works have
been and remain incredibly influential and important to the field of
integrable systems. Francesco himself is still developing the mathematics
associated with this area. (Just look over the 200+ papers on
MathSciNet that he has written!) His contributions will surely influence the
next generation of people working in this area,  as they have with the
current generation. Of Francesco's work, the references 
that are most directly
relevant to the topics discussed in the present paper are
\cite{calogero,CaFr1996}.

\section{Derivation of the functional equation}
\setcounter{equation}{0}

Let us start from any equation (\ref{ipde}) 
in the pulson $b$-family. 
The key is to introduce, 
for any kernel  $g$, 
the operator 
\beq \label{skewop} 
\hat{B}=-(b m \partial_x+ m_x)\hat{G} 
(b m \partial_x +(b-1)m_x), 
\eeq 
where the operator $\hat{G}$ acts according to 
$$ 
\hat{G}f(x)=G*f=\int_{-\infty}^\infty G(x-y)f(y)\, dy, 
$$ 
with $G$ being the antisymmetric kernel (\ref{anti}).   
Thus $\hat{B}$ is skew-symmetric and for $b\neq 1$ satisfies 
$$
m_t=\hat{B} \frac{\delta H}{\delta m}\, 
\,,\qquad H=\frac{1}{(b-1)}\int_{-\infty}^\infty m\, dx. 
$$ 

Now we can calculate the bracket defined by the 
operator $\hat{B}$, namely  
$$ 
\{ F, G\}=\int_{-\infty}^\infty \frac{\delta F}{\delta m} 
\hat{B}\frac{\delta G}{\delta m}\, dx, 
$$ 
which 
gives  
$$
 \{ m(x),m(y) \} =\Big( 
G(x-y)m_x(x)m_x(y)+bG'(x-y)(m(x)m_x(y)-m(y)m_x(x)) 
$$ 
\begin{equation}
         - b^2 G''(x-y)m(x)m(y)\Big) . 
\label{eq:pb} 
\end{equation} 
Although (\ref{eq:pb}) is antisymmetric (since $G$ is odd), 
it only defines a  Poisson bracket if the Jacobi identity is 
satisfied, i.e. 
\begin{equation} 
\{ \{ m(x),m(y)\} , m(z)\} + {\bf cyclic} =0. 
\label{eq:jac} 
\end{equation} 
To calculate (\ref{eq:jac}), we evaluate 
$$ 
\frac{\delta}{\delta m(s)} \{ m(x),m(y) \} = 
\Big( 
G(x-y)[\delta '(x-s)m_x(y)+\delta '(y-s)m_x(x)] 
$$ 
$$  
+bG'(x-y)[\delta (x-s)m_x(y)-\delta (y-s)m_x(x) 
+\delta '(y-s)m(x)-\delta '(x-s)m(y)]  
$$ 
$$  
- b^2 G''(x-y)[\delta (x-s)m(y)+\delta (y-s)m(x)]  
\Big) ,
$$ 
and then substitute into 
$$ 
\{ \{ m(x),m(y)\} , m(z)\}=\int_{-\infty}^\infty 
\frac{\delta}{\delta m(s)} \{ m(x),m(y) \} \, 
\hat{B}(s) \delta(z-s) 
\, ds 
$$ 
(where $\hat{B}(s)$ indicates the operator acting in 
the independent variable $s$). 

In the Jacobi identity (\ref{eq:jac}) many terms cancel to 
leave only terms cubic in $m$ and $m_x$, whose  
coefficients 
must each vanish if the identity holds  for arbitrary $m$.
This leads to functional equations for the odd function $G$, 
but remarkably (apart from overall prefactors) these 
coefficients  are independent of the parameter $b$.  
Thus, setting 
$$
\alpha=x-y, \quad \beta=y-z, \quad \gamma=z-x, 
\qquad \mathrm{whence} \qquad \alpha+\beta+\gamma=0,  
$$ 
we find the coefficient 
of $m_x(x)m_x(y)m_x(z)$ yields the functional equation 
(\ref{eq:fn1}). 
The coefficient of $m(x)m_x(y)m_x(z)$ gives 
\begin{equation}  
G'(\beta)(G'(\alpha)-G'(\gamma))- 
G''(\gamma)(G(\alpha)+G(\beta)) 
+G''(\alpha)(G(\beta)+G(\gamma))=0,     
\label{eq:fn2} 
\end{equation} 
and there are two other identities like (\ref{eq:fn2})
obtained by cyclic 
permutations of $x,y,z$ which induce cyclic permutations 
of $\alpha,\beta,\gamma$ 
(with the three identities summing to zero). 
Similarly the coefficient of $m_x(x)m(y)m(z)$ yields 
\begin{equation}  
G'''(\beta)(G(\alpha)+G(\gamma))+ 
G''(\gamma)(G'(\alpha)-G'(\beta)) 
-G''(\alpha)(G'(\beta)-G'(\gamma))=0,     
\label{eq:fn3} 
\end{equation}   
and there are two other identities obtained by cyclic 
permutation of $\alpha,\beta,\gamma$ 
in (\ref{eq:fn3}). Finally the 
coefficient of $m(x)m(y)m(z)$ requires 
\begin{equation} 
G'''(\alpha)(G'(\beta)-G'(\gamma)) 
+G'''(\beta)(G'(\gamma)-G'(\alpha))+ 
G'''(\gamma)(G'(\alpha)-G'(\beta))=0. 
\label{eq:fn4} 
\end{equation}  

There are several things to observe about these 
functional equations. Firstly, (\ref{eq:fn2}) and 
(\ref{eq:fn3}) are direct consequences of 
(\ref{eq:fn1}), obtained 
by differentiating. The final functional equation 
(\ref{eq:fn4}) is also a  
consequence of the rest. Hence the Jacobi identity 
is satisfied by the operator $\hat{B}$ if and only if 
the antisymmetric kernel $G$ satisfies the functional 
equation (\ref{eq:fn1}). We can prove the following: 

\noindent {\bf Proposition.} {\it Suppose that $G(x)$ is an odd solution 
of the functional equation (\ref{eq:fn1})  
with continuous second derivative $G''(x)$ for all $x\in\mathbb{R}\backslash \{ 0 \} $, 
and suppose $\mathrm{lim}_{x\to 0\pm}G''(x)$ is finite. 
Then either $G(x)=A\, x$ or $G(x)=A\,\mathrm{sgn}(x)\Big(1-e^{-B|x|}\Big)$ 
for some constants $A,B$. } 

\noindent {\bf Sketch of proof.}  If $G$ is odd and continous 
second differentiable then from the limit $\alpha\to 0$ in (\ref{eq:fn1}) 
we see that $G(0)=0$, while differentiating (\ref{eq:fn1}) 
with respect to $\alpha$ yields (\ref{eq:fn2}). Taking the  
limit $\alpha\to 0$ in (\ref{eq:fn2}) with $\beta$ fixed gives the 
differential equation 
$$ 
G''(\beta )G(\beta )-G'(\beta )^2+G'(0 )G'(\beta ) =0, 
$$ 
which integrates to 
$$ 
G'(\beta )=G'(0 )+K_{\pm}G(\beta ), 
$$ 
where the integration constants $K_\pm$ can be different for 
$\beta >0$ and $\beta <0$. This is easily integrated to give the 
solution for $G$. 

\noindent {\bf Remarks.} A more detailed derivation of the general 
solution (not necessarily odd), with 
weaker assumptions, appears as an Appendix. Here we should 
observe that if the parameter $B\to \infty$ with $A$ fixed then the 
distributional  solution 
$$G(x)=A\,\mathrm{sgn}(x), \qquad g(x)=2A\,\delta (x)$$ 
results, 
in which case for $A=1$  
(\ref{ipde}) becomes the Riemann shock equation 
$$ 
m_t+2(b+1)mm_x=0, $$ 
while when $B\to 0$ with $A\sim 1/B$ the linear solution 
$G(x)=Ax$ arises. The latter gives only the trivial linear 
PDE 
$$ 
m_t+Km_x=0, \qquad K=A\int m(y) \, dy. 
$$


\section{Restricting the bracket to the pulsons} 
\setcounter{equation}{0}

To calculate the corresponding bracket for the restriction to 
the finite-dimensional pulson submanifold,   
we substitute $m=\sum p_j \delta (x-q_j)$ into both sides 
of (\ref{eq:pb}), which gives 
$$ 
\sum_{j,k}\Big( \{ p_j,p_k \}\delta (x-q_j) \delta (y-q_k) 
-\{ p_j,q_k \} p_k \delta (x-q_j) \delta '(y-q_k) 
$$ 
$$ 
-\{ p_j,q_k \} p_j \delta '(x-q_j) \delta (y-q_k)   
+ \{ q_j,q_k \}p_jp_k \delta '(x-q_j) \delta '(y-q_k)\Big) = 
$$ 
$$ 
\sum_{j,k}p_jp_k\Big(G(x-y)\delta '(x-q_j) \delta '(y-q_k) 
+bG'(x-y)(\delta (x-q_j) \delta '(y-q_k) - 
\delta '(x-q_j) \delta (y-q_k) ) 
$$ 
$$ 
-b^2G''(x-y)\delta (x-q_j) \delta (y-q_k)\Big). 
$$   
We fix $\epsilon$ with 
$0<\epsilon <\min_{j\neq k}|q_j-q_k|$ 
and then introduce  the indicator functions 
$$ 
\begin{array}{cccl} 
I_j(x) & =& 1,\,\, & x\in [q_j-\epsilon,q_j+\epsilon ], \\ 
       & =& 0, & otherwise, 
\end{array} 
$$ 
and 
$$ 
J_j(x)=(x-q_j)I_j(x). 
$$ 
Integrating the Poisson bracket relation against  
suitable products of 
indicator functions centred around $x=q_j$, $y=q_k$, 
such as $I_j(x)I_k(y)$ and $I_j(x)J_k(y)$ etc. we find  
$$ 
\{ p_j,p_k \}=-(b-1)^2G''(q_j-q_k)p_jp_k, \quad  
\{ q_j,p_k \}=(b-1) p_k G'(q_j-q_k),  
$$ 
\begin{equation} 
\{ q_j,q_k \}= G(q_j-q_k).  
\label{eq:fibracket} 
\end{equation} 
The brackets (\ref{eq:fibracket}) 
generate the pulson $b$-equations by taking the 
restriction of $H$ to the finite submanifold  
i.e. 
$$ 
H=\frac{1}{(b-1)}\int_{-\infty}^\infty \sum_k p_k\delta(x-q_k)\, dx  
=\frac{1}{(b-1)} \sum_k p_k. 
$$ 
Thus we find that the pulson dynamical system (\ref{eq:peakdyn}) becomes  
$$ 
\frac{dp_j}{dt}=\{ p_j,H\}, \qquad \frac{dq_j}{dt}=\{ q_j,H\} 
$$ 
(with  $G'=g$). 
But (\ref{eq:fibracket}) 
are only Poisson brackets if the Jacobi 
identity is satisfied. 
For the Jacobi identity of the finite-dimensional 
brackets we require both  
\begin{equation} 
\{\{ q_j,q_k \} ,p_l\}+{\bf cyclic}=0 
=\{\{ p_j,p_k\} ,q_l\}+{\bf cyclic}. 
\label{eq:fijac} 
\end{equation} 
for all $j,k,l$, and similarly 
\begin{equation}
\{\{ q_j,q_k \} ,q_l\}+{\bf cyclic}=0
=\{\{ p_j,p_k\} ,p_l\}+{\bf cyclic}.
\label{eq:fijac2}
\end{equation}  
The necessary conditions resulting from 
(\ref{eq:fijac}) and \ref{eq:fijac2})  
are the functional  
equations (\ref{eq:fn1}),  
(\ref{eq:fn2}),  
(\ref{eq:fn3}), (\ref{eq:fn4})  and their cyclic permutations. 
Therefore if $G$ satisfies the functional equation 
(\ref{eq:fn1}) then this guarantees 
that (\ref{eq:fibracket}) is a finite-dimensional Poisson bracket.  

\section{Conclusions} 

The functional equation (\ref{eq:fn1}) provides an elegant way 
to pick out the peakon equations (\ref{eq:bfamily}) 
from the whole class of IPDEs (\ref{ipde}). 
However, it is interesting that the compacton type solution 
$G(x)=x(1-|x|/2)H(1-x)H(1+x)$ does not satisfy the equation. 
The compactons  are solitons with compact support which 
appear in the Hunter-Saxton \cite{olveros} or Vakhnenko equations 
\cite{parkes3, vak}. In that case $G$ is formally a Green's 
function for $\partial_x^{-3}$, and the operator (\ref{skewop}) 
was formally identified as satisfying the Jacobi identity 
in \cite{wanghone}. However, in \cite{wanghone} we noted that 
the Vakhnenko equation is the short wave limit of 
the Degasperis-Procesi equation, and so there may be problems 
with defining the boundary conditions correctly.  

\noindent {\bf Acknowledgements.} We are grateful to Martin Staley for
providing the Figures. Both authors would like 
to thank Toni Degasperis for introducing us to the Degasperis-Procesi 
equation. We are also grateful for the hospitality of 
the Isaac Newton Institute, where we first began working on $b=3$ in 
2001. AH thanks Harry Braden and John Byatt-Smith for providing the 
general solution of the functional equation, and is grateful to 
Andrey Leznov for suggesting the simplest approach when $G$ is odd. 
DH and AH also thank Simonetta Abenda and Tamara Grava for    
inviting us to the meeting {\it Analytic and Geometric theory 
of the Camassa-Holm equation and Integrable Systems} in Bologna, in 
September 2004, where we decided to finish writing up these results. 
DH thanks AH for taking the lead in finishing our project and
finally transforming it from a long string of email exchanges into a
finished product. We are both grateful to Francesco Calogero for many
kindnesses and inspiring discussions over the years of our association
together.

\newtheorem{thm}{Theorem}[section]

\section*{Appendix: Solution of the Functional Equation 
{\it by H.W. Braden and J.G. Byatt-Smith}}
\setcounter{equation}{0}


Here we shall solve the functional equation
\begin{equation}\label{fnl}
G'(x)\left[ G(y)+G(z)\right] +G'(y)\left[ G(z)+G(x)\right]
+G'(z)\left[ G(x)+G(y)\right] =0,
\end{equation}
where $x+y+z=0$ and $G(x)$ is differentiable with continuous first
derivative. 
We prove the following
theorem.
\begin{thm} Suppose $G(x)$ is a solution of equation (\ref{fnl}) that
is differentiable  with continuous first derivative. If $G(x)$
possesses (possibly different) power series expansions in the
intervals $(-\epsilon,0)$ and $(0,\epsilon)$ (for $\epsilon>0$)
then, up to the invariance of the functional equation
\begin{equation}\label{fnlinv}
G(x)\rightarrow a\, G(Ax),
\end{equation}
$G(x)$ is one of
\begin{equation}\label{fnlsola}
(i) \qquad \qquad G(x) =\frac{1}{c}\left[ e^{Ax}+e^{-Ax}+1\right],
\end{equation}
which for $A=0$ gives the constant solution.

\begin{equation}\label{fnlsolb}
(ii) \qquad \qquad G(x) = \sin(x)\sin(x-\frac{\pi}{3}),
\end{equation}
which has the scaling limit $G(x)=\alpha\, x$.

\begin{equation}\label{fnlsolc}
(iii) \qquad \qquad G(x) ={\rm sgn}(x)\left[ 1-e^{-|x|}\right].
\end{equation}

\end{thm}

We begin with some preliminary observations.
\begin{enumerate}
\item  Equation (\ref{fnl}) is satisfied for $G$ a constant.
Henceforth we assume that $G$ does not vanish identically.

\item $G(0)\,G'(0)=0$. This is seen by setting $0=x=y=z$ in
(\ref{fnl}).

\item Suppose that both $G(0)=G'(0)=0$. Then with $x=0$ and $z=-y$
we find that
$$G'(y)\,G(-y)+G'(-y)\,G(y)=0.$$
Thus $G(-y)/G(y)=\lambda$, a constant. Then $\lambda=\pm1$, and
therefore $G$ is either even or odd.

 \item With $x=y$ we obtain
\begin{equation}\label{fnlredpr}
G(-2x)\,G'(x)+G'(x)\,G(x)+G'(-2x)\,G(x)=0, \end{equation}
and so
$$\frac{d}{dx}\left[ \frac{G(-2x)}{G^{2}(x)}+\frac{2}{G(x)}\right]
=0.$$ Therefore, on an interval $I$ for which $G(x)\neq 0$ we have
\begin{equation}\label{fnlred}
    G(-2x) + 2 G(x) = c_{I}\, G^{2}(x)
\end{equation}
It is possible for the constant $c_{I}$ to vary from interval to
interval. This means that the function can have different power
series expansions in different intervals. There are solutions
possessing different power series on different intervals. For
example (\ref{fnlsolc}) satisfies (\ref{fnlred}) with $c=1$ for
$x>0$ and $ c =-1$ for $x<0$. (We will see how this arises below.)

Thus solutions of equation (\ref{fnlred}) contain those of
(\ref{fnl}).

\item From (\ref{fnl}) we deduce that if $G'$ is differentiable at
the point $y$ and $G(y)+G(-x-y)\ne0$, then $G'$ is also
differentiable at the point $x$. Thus if $G'$ is differentiable
for $0<|x_0-y|<\epsilon$, but not at $x_0$ we have
$$G(y)+G(-x_0-y)=0.$$
Using the continuity of $G$ we have
$$G(x_0)+G(-2x_0)=0.$$
We deduce from (\ref{fnlredpr}) that at such a point
$$G(x_0)\,G'(-2 x_0)=0.$$

\item We record
\begin{equation}\label{fnlredrec}
\begin{split}
 -G'(-2x)&=c_I\,G(x)\,G'(x)-G'(x)\\
2\,G''(-2x)&=c_I\,\left[ G'(x)^{ 2}+G(x)\,G''(x)\right] -G''(x)\\
-4\,G'''(-2x)&=c_I\,\left[ 3\,G'(x)\,G''(x)+G(x)\,G'''(x)\right]
-G'''(x)\\
8\,G\sp{(iv)}(-2x)&=c_I\,\left[ 4\,G'(x)\,G'''(x)+3\,G''(x)^{
2}+G(x)\,G\sp{(iv)}(x) \right] -G\sp{(iv)}(x)
\end{split}
\end{equation}

\end{enumerate}

\subsection*{Proof of the Theorem}
The strategy of our proof is the following. We have observed that
the solutions of equation (\ref{fnlred}) contain those of
(\ref{fnl}). We will first find the solutions of equation
(\ref{fnlred}) satisfying the conditions of the theorem, and then
verify that these in fact solve (\ref{fnl}).

Observe from (\ref{fnlred}) that if $G(x)$ is defined in some
neighbourhood of $x=0$ it may be extended to the whole of the real
line. We will assume that $G(x)$ has a power series expansion for
$0<|x|<\epsilon$, possibly different for $x$ positive and
negative. To distinguish between the possibly different power
series for $G$ set
\begin{equation}
\begin{array}{lc}
g(x)=G(x),&{\rm if}\quad x>0,\\
f(x)=G(-x),&{\rm if}\quad x<0.
\end{array}
\end{equation}
Then from (\ref{fnlred})
\begin{align}
f(2x)&=c_{+}\, g(x)^2-2\,g(x),\label{fnlredf}\\
g(2x)&=c_{-}\, f(x)^2-2\,f(x).\label{fnlredg}
\end{align}
In particular, this means that
\begin{equation}
f(4x)= c_{+}\,\left[c_{-}\,
f(x)^2-2\,f(x)\right]^2-2\,\left[c_{-}\,
f(x)^2-2\,f(x)\right].\label{fnlredfx}
\end{equation}

Our assumptions on the continuity of $G$ and $G'$ mean that
$f(0)=g(0)$ and $f'(0)=g'(0)$. We also note from (\ref{fnlred})
that
\begin{equation}
3\,f(0)=c_{-}\, f(0)^2, \qquad 3\,g(0)= c_{+}\, g(0)^2.
\label{fnlredbc}
\end{equation}
Now we two distinct cases, as follows:   

\noindent {\bf Case (1)} If $f(0)=g(0)\ne0$, then from (\ref{fnlredbc}) we deduce
that $c_-=3/f(0)=3/g(0)=c_+$. Therefore $f(x)$ and $g(x)$ satisfy
the same recursion with the same initial conditions, whence
$f(x)=g(x)$. Together with say (\ref{fnlredf}) this yields a
recursion for the power series coefficients of the function (see
for example (\ref{fnlredrec})) which may be solved to give
\begin{equation}
G(x) =\frac{1}{c}\left[ e^{Ax}+e^{-Ax}+1\right].
\end{equation}
With $A=0$  this gives $f(x)=3/c$, the constant solution noted
earlier.


\noindent {\bf Case (2)} If $f(0)=g(0)=0$ then there are two 
possibilities:  

\noindent {\bf (a)} 
If $f'(0)=g'(0)=0$ then from our earlier observation we have
that $f(x)=\pm g(x)$. Substituting the power series
$f(x)=c_2\,x^2+\dots$ into (\ref{fnlredf}) and employing this
observation shows $f(x)=0$, and consequently $$G(x)=0.$$

\noindent {\bf (b)}  
Thus we may suppose $f'(0)=g'(0)\ne0$. Now from
(\ref{fnlredrec}) we find that in this case
\begin{align*}
3\,f''(0)&=(2\,c_+ -c_-)f'(0)^2,\\
3\,g''(0)&=(2\,c_- -c_+)g'(0)^2.
\end{align*}
\begin{enumerate}
\item If the second derivatives agree then $c_+=c_-$ and again
$f(x)=g(x)$. Solving the recursion yields
\begin{equation}
G(x) = \sin(x)\sin(x-\frac{\pi}{3}) =-\frac{\sqrt{3}}{2}x+\frac{%
x^{2}}{2}+\frac{x^{3}}{3}-\frac{x^{4}}{6}+\dots
\end{equation}
This satisfies $G(-2x) +2G(x)=4G^{2}(x)$; that is $c=4$. A scaling
limit of this solution is $$G(x)=\alpha\, x,$$ which corresponds
to $c=0$. (This is obtained by
 $G(x)\rightarrow a\,G(x)$, under which
$ c\rightarrow \frac{c}{a}$.)

\item The final possibility corresponds to $f$ and $g$ having
different second derivatives at $x=0$. This corresponds to $G'$
not being differentiable at this point. Our earlier observation
then yields $G(y)+G(-y)=0$. That is $G$ is odd, whence
$f(x)=-g(x)$ and consequently $c_+=-c_-$. Solving the recursion
yields
\begin{equation}
G(x) ={\rm sgn}(x)\left[ 1-e^{-|x|}\right].
\end{equation}
\end{enumerate}


At this stage we have established the Theorem. 


\noindent{\bf Remark.} In \cite{BBS} an alternative
method is described 
for investigating functional equations like (\ref{fnl}). It
is instructive to compare this with the approach above. Upon
setting $y=x-t$ and $z=t-2x$ we expand (\ref{fnl}) in $t$: this
leads to an infinite set of differential equations that $G$ must
satisfy. The term independent of $t$ yields (\ref{fnlredpr}) and
the equation from the order $t$ term is satisfied identically as a
consequence of (\ref{fnlred}). If we substitute (\ref{fnlred}),
which involves the arbitrary constant $c$, into the order $t^2$
and $t^3$ equations resulting from the expansion we can eliminate
the constant $c$ between them. This yields the following simple
result
$$0=g(x)\,g''(x)\,\left[g^{(iv)}\,g'(x)-g''(x)\,g'''(x)\right],$$
from which we see that either $g$ vanishes, is a linear function,
or satisfies the fourth order differential equation given by the
last term. The general solution of this latter equation is
\begin{equation}\label{fnlfths}
G(x)=a\,e^{Ax}+b\,e^{-Ax}+d.
\end{equation}
As we have obtained a necessary condition for a function to
satisfy (\ref{fnl}) we may now substitute (\ref{fnlfths}) into
either the original equation or (\ref{fnlred}), placing
restrictions on the constants. This leads to the solutions above.

For example, we may express the functions vanishing at $x=0$ by
\begin{equation}
f(x)=\frac{\left(1-u(x)\right)\left(u(x)-\beta\right)}{u(x)\left(1+\beta\right)c_-}
\end{equation}
where $u(x)=e\sp{x}$ and
$c_+/c_-=-(1+\beta )^2/( 1+\beta^2)$.
Then from (\ref{fnlredfx})
\begin{equation}
f(4x)=\frac{\left(1-u(x)^4\right)\left(u(x)^4-\beta^4\right)}{u(x)^4\left(1+\beta^2\right)\left(1+\beta\right)^2c_-}
=\frac{\left(1-u(4x)\right)\left(u(4x)-\beta\right)}{u(4x)\left(1+\beta\right)c_-}
\end{equation}
Solving this leads to $\beta=0$ and the solution (\ref{fnlsolc}),
or to $\beta=\omega$, $\omega^2$, where $\omega$ is a nontrivial
cube-root of unity and the solution (\ref{fnlsolb}). (Here we have
used that $u(x)=e\sp{x}$. However, when $\beta=0$, $\omega$, or
$\omega^2$ and $u(0)=1$ we may in fact deduce that
$u(x)=e\sp{x}$.)


\begin{thebibliography}{99}
\small

\bibitem{Acz}J. Acz\'el, {\it Lectures on functional equations and their  
applications}, Academic Press, New York 1966.  

\bibitem{beals}R. Beals, D. Sattinger, J. Szmigielski, 
{\it Multipeakons and the classical moment problem},  
Adv. Math.  154  (2000) 229--257. 

\bibitem{beals2}R. Beals, D. Sattinger, and J.
Szmigielski, 
{\it Multi-peakons and a theorem of Stieltjes},  Inverse Problems 
15 (1999) 
L1-L4. 

\bibitem{beals3} 
R. Beals, D. Sattinger, and J. Szmigielski, 
{\it Peakon-antipeakon interaction},  
J. Nonlinear Math. Phys.  8 (2001) Suppl. 23-27.

\bibitem{BBS}H.W. Braden and J.G.B. Byatt-Smith,  
{\it On a Functional Differential Equation of Determinantal Type},  
Bull. Lond. Math. Soc.  31 (1999) 463-470.

\bibitem{ch}R.~Camassa and D.D.~Holm,
{\it An integrable shallow water equation with peaked solitons}, 
Phys. Rev. Lett. 71 (1993) 1661-1664.

\bibitem{ch2}R.~Camassa, D.D.~Holm and J.M.~Hyman,
{\it A New Integrable Shallow Water Equation},
Advances in Applied Mechanics 31 (1994) 1-33.

\bibitem{calogero}F.~Calogero, {\it Classical many-body problems
amenable to exact treatments,} Lecture Notes in Physics Monograph
66, Springer--Verlag (2001). 

\bibitem{CaFr1996}
F.~Calogero and J.-P.~Francoise, 
{\it A completely integrable Hamiltonian system}, 
J. Math. Phys. 37 (1996) 2863-2871. 
 
\bibitem{constantin} 
A. Constantin and H. P. McKean,  
{\it A shallow water equation on the circle
},  
Comm. Pure Appl. Math.  52 (1999) 949-982. 

\bibitem{constantin3} 
A. Constantin and L. Molinet, {\it Global 
weak solutions for a shallow water equation}, Comm. Math.
Phys. 211 
(2000) 45-61. 
 

\bibitem{constantin2}
A. Constantin, 
{\it On the scattering problem for the Camassa-Holm equation},  
R. Soc. Lond. Proc. Ser. A Math. Phys. Eng. Sci.  457  (2001) 
953-970. 

\bibitem{constantin4}
A. Constantin and W. Strauss, {\it Stability of peakons},  
Comm. Pure Appl. Math.  53 (2000) 603--610. 

\bibitem{constantin5} 
A. Constantin 
and L. Molinet, {\it Orbital stability of solitary waves for a shallow 
water equation},  Physica D  157  (2001) 75-89.

\bibitem{constantin6}
A. Constantin and B. Kolev, {\it On the geometric approach to the 
motion of inertial mechanical systems},  J. Phys. A 35 (2002)
R51-R79. 

\bibitem{constantin7}
A. Constantin and B. Kolev, 
{\it Geodesic flow on the diffeomorphism 
group of the circle},  Comment. Math. Helv.  78 (2003)
787-804.

 

\bibitem{dega}A. Degasperis and M. Procesi,
{\it Asymptotic integrability}, in
{\it Symmetry and Perturbation Theory} (eds. A. Degasperis and G.
Gaeta), World Scientific (1999) 23-37.

\bibitem{needs}A.~Degasperis, A.N.W.~Hone and D.D.~Holm,
{\it A New Integrable Equation with Peakon
Solutions}, 
Theoret. Math. 
Phys. 133 (2002) 1461-1472. 

\bibitem{galli}A.~Degasperis, A.N.W.~Hone and D.D.~Holm,
{\it Integrable and non-integrable equations with peakons},
in {\it Nonlinear Physics: Theory and Experiment II}, eds. M.J.~Ablowitz, 
M.~Boiti, F.~Pempinelli \& B.~Prinari,  
World Scientific (2003) 37-43; {\tt nlin.SI/0209008}

\bibitem{dgh1}H.R.~Dullin, G.A.~Gottwald and D.D.~Holm, 
{\it Camassa-Holm, Korteweg--de Vries-5 and other 
asymptotically equivalent equations for shallow water}, 
Fluid Dynamics Research 33 (2003) 73-95. 

\bibitem{dgh2}H.R.~Dullin, G.A.~Gottwald and D.D.~Holm, 
{\it On asymptotically equivalent shallow water wave equations}, 
Physica D 190 (2004) 1-14. 

\bibitem{fringer}O.~Fringer and D.D.~Holm,
{\it Integrable vs. nonintegrable geodesic soliton behaviour}, 
Physica D 150 (2001) 237-263.

\bibitem{ff}B.~Fuchssteiner and A.S.~Fokas, 
{\it Symplectic structures, their B\"acklund transformations, 
and hereditary symmetries},  
Physica D 4 (1981) 47.



\bibitem{hs}D.D.~Holm and M.F.~Staley,
{\it Wave Structures and Nonlinear Balances in a Family of
1+1 Evolutionary PDEs}, 
SIAM J. Appl. Dyn. Syst. 2 (2003) 323-380. 

\bibitem{hoho}D.D.~Holm and A.N.W.~Hone,
{\it Nonintegrability of a fifth order equation
with integrable two-body dynamics}, 
Theoret. Math. Phys. 137 (2003) 1459-1471. 

\bibitem{homa}D.D.~Holm and  J.E.~Marsden, 
{\it Momentum Maps and Measure-valued Solutions (Peakons, 
Filaments and Sheets) for the EPDiff Equation}, 
{\tt nlin.CD/0312048} 

\bibitem{hone1}A.N.W.~Hone,
{\it The associated Camassa-Holm equation and the KdV equation},
J. Phys. A 32 (1999) L307-L314.


\bibitem{hone3}A.N.W.~Hone,
{\it Reciprocal link for 2+1-dimensional Extensions of
Shallow Water Equations},
Applied Mathematics Letters 13 (2000) 37-42.

\bibitem{wanghone}A.N.W.~Hone and Jing Ping Wang, 
{\it Prolongation algebras and Hamiltonian operators for
peakon equations},  
Inverse Problems 19 (2003) 129-145.   
 
\bibitem{pain}A.N.W.~Hone,  
{\it Painlev\a'{e} tests, singularity structure and integrability},
in {\it What is Integrability?}, ed. A.V.~Mikhailov, 
Princeton University Press, to appear;  
preprint (2003) UKC/IMS/03/33.  

\bibitem{johnson}R. S. Johnson, 
{\it Camassa-Holm, Korteweg-de Vries and related models for water
waves}, 
J. Fluid Mech.  455 (2002)  63-82.  


\bibitem{kraenkel}R.A.~Kraenkel and A.~Zenchuk, 
{\it Two-dimensional integrable generalization of the 
Camassa-Holm equation} 
Phys. Lett. A 260 (1999) 218-224. 

\bibitem{mik}A.V.~Mikhailov 
and V.S.~Novikov, {\it Perturbative symmetry approach},  
J. Phys. A 35 (2002) 4775-4790.

\bibitem{misiolek} 
G. Misiolek, {\it A shallow water equation as a 
geodesic flow on the Bott-Virasoro group},  J. Geom. Phys. 24
(1998)  
203-208. 

\bibitem{olver}P.J.~Olver, {\it Applications of Lie Groups
to Differential Equations}, 2nd edition,
Springer-Verlag (1993).

\bibitem{olveros}P.J.~Olver and P.~Rosenau,
{\it Tri-Hamiltonian Duality Between Solitons and Compactons},
Phys. Rev. E 53 (1996) 1900-1906.



\bibitem{parkes3}E.J.~Parkes and V.O.~Vakhnenko,
{\it The calculation of multi-soltion solutions 
of the Vakhnenko equation by the inverse scattering method}, 
Chaos, Solitons and Fractals 13 (2002) 1819-1826.





\bibitem{vak}V.O.~Vakhnenko, {\it Solitons in a nonlinear model 
medium}, J. Phys. A 25 (1992) 4181-4187.

\end{thebibliography}
\end{document}